\documentstyle[aps]{revtex}
\input epsf

\begin{document}
\title{A scalable, tunable qubit, based on a clean DND or grain boundary
 D-D junction}
\author{Alexandre M. Zagoskin\thanks{%
email: zagoskin@physics.ubc.ca}}
\address{Physics and Astronomy Dept., The University of British\\
Columbia, 6224 Agricultural Rd, \\
Vancouver, B.C., V6T 1Z1 Canada}
\maketitle

\begin{abstract}
{Unique properties of a  ballistic DND  or grain boundary D-D junction,
including doubly degenerate ground state with tunable potential barrier
between the "up" and "down" states and non-quantized spontaneous magnetic
flux,
make it a good candidate for a solid state qubit. The role of
quantum "spin" variable is played by the sign of equilibrium superconducting
phase difference on the junction, which is revealed in the direction of
spontaneous supercurrent flow in equilibrium.
 Possibilities of design-specific simultaneous
operations with several integrated qubits are discussed.}
\end{abstract}

The pronounced shift from "software" to "hardware" in theoretical research
on quantum computing (QC) \cite{DiV,DiVLoss,Schon,Ioffe}   is a good measure 
of growing
confidence that QC can be realized on practically interesting scale (of at
least $10^3$ qubits). Though first experimental realizations of QC used such
technologies as NMR  and ion trapping, the
problem of scalability for such approaches still looks formidable. Therefore
much effort is directed at search for a practical solid state qubit (SSQ), with
natural candidates being such mesoscopic devices as quantum dots
 \cite{DiV,DiVLoss}
,
mesoscopic Josephson junctions\cite{Schon,Ioffe} and superconducting
single-electron transistors (parity switches) \cite{Schon}. The evident
advantage of a SSQ is scalability, where all the potential of
existing solid state technologies could be used\cite{DiV}, while
among the problems the main are (1) to achieve quantum beatings between
distinguishable states of a single qubit, (2) to prevent loss of coherence during calculations,
and (3) to minimize statistical dispersion
of the properties of individual qubits.

The problems (1) and (2) pose specific difficulties in an SSQ 
due to huge number of
degrees of freedom coupled
to it, and to necessity to fine-tune two states of a 
mesoscopic system  chosen as working ones 
to a resonance.
  
A possibility to circumvent these obstacles is
presented by Josephson systems with d-wave cuprates, which violate
time-reversal symmetry and as a result have doubly degenerate groundstate
\cite{Sigrist,Huck} with a potential for quantum beatings (macroscopic
quantum tunneling), or at least quantum noise \cite{ZagoskinOshikawa}. Ioffe
et al. recently incorporated this property in their ''quiet qubit'' design\cite{Ioffe},
which uses tunneling (SID) or dirty SND junctions with equilibrium
phase difference $\varphi _0=\pm \frac{\pi}{2}$.

Let us consider a device shown in Fig.1a. Its main part is
 a {\em clean}
mesoscopic  D-D junction (i.e. ballistic DND or D-(grain boundary)-D junction).
Though we will concentrate on the DND qubit design, the same considerations are 
applicable, mutatis mutandis, to high quality
grain-boundary D-D junctions, where recently an analogous
current-phase dependence was observed\cite{Ilichev}. 
The grain boundary region where the superconducting gap is suppressed,
is naturally modeled by a normal conductor with the same 
lattice parameters and chemical potential as in the superconducting banks,
which only enhances the amplitude of purely Andreev scattering in the system.

The terminal B of the junction is formed by a massive d-wave superconductor;
in a multiple-qubits  system, they will all use it as a common "bus" bar.
The terminal A is small enough to allow - when isolated -
quantum phase fluctuations. It is essentially
 the sign of the superconducting phase
difference $\varphi$ between the terminals A and B that plays the role of "spin 
variable" of quantum computing. The collapse of the wave function is achieved
by connecting the terminal A with the external source
of electrons ("ground"), thus blocking the phase fluctuations due to
phase-number uncertainty relation\cite{Tinkham}. So far, the
best way to do this is presented by using a "parity key", PK (superconducting
single-electron transistor)\cite{Gisselfalt,Ktototam}, which  only
passes Cooper pairs, and only at a certain gate voltage $V_g$.  
Other parity keys, with different parameters, are used to link adjacent qubits,
allowing for controllable entanglement (Fig.1b). Such an architecture
allows reasonably easy way to integrate a large number of qubits in a
1D or 2D matrix. We will see that it also provides a natural
way of 
preparing  all qubits on the same bus in the same initial quantum state,
 thus facilitating the implementation of error correction
algorithms.

The readout of the state of a qubit is simplified by the presence of 
small spontaneous, non-dissipative currents 
and spontaneous fluxes (of order $10^{-2}-10^{-3}\Phi_0$ depending on the
setup) concentrated
 in the central  part of the DND junction\cite{Huck,ZagoskinOshikawa},
  which have {\em opposite directions} 
 in two degenerate equilibrium states.
 While too small to lead to
unwanted inductive coupling between the qubits or 
decoherence, they can be still used to read out the state of the qubit once it
was collapsed in one of the states with $\pm \varphi _0$, e.g. using the magnetic force microscope
tip M (which is removed during the computations). Collapsing and
reading processes are thus time separated, and the computation results
will be automatically preserved for the time limited only by the thermal
fluctuations.

A necessary condition  for a qubit to work is $
t_{t(unneling)} < t_{g(ate\:\:application)} < t_{d(ecoherence)}$.
Here a clean DND junction has a very important advantage following from
the fact that 
  the absolute value of equilibrium phase
difference, $|\varphi _0|$, can now vary from $0$ to $\pi$  
depending on the angle $\Omega$ between the crystal axes of the d-wave
superconductors and the ND boundary \cite{Zagoskin}. 
In practice, since
the lattice structure of the d-wave superconductors allows only a limited
set of easy cleavage directions, e.g. $\left( 010\right) $ and $(100)$,  the
equilibrium phase can be varied by preparing a steplike  ND interface, with
the equilibrium phase determined
  by the
relative weight of
Andreev zero- and $\pi -$levels (coupled to the lobes of d-wave order parameter 
in A and B with
the same or opposite sign respectively),   produced by such an arrangement. 
Since the shape
of the effective potential barrier between states with $\pm\varphi _0$
depends on $\varphi_0$, it can now be varied.  Therefore the tunneling rate can be chosen in 
exponentially wide limits to
achieve the optimal performance.   
(In an SND junction the equilibrium 
phase can also be varied, but it cannot be made less than 
$ \pi(\sqrt{2}-1)/\sqrt{2}$)\cite{Zagoskin}.)
Besides, this allows to fix the working
interval of the device to $|\varphi|\leq\pi$, since due to exponential dependence of the
tunneling amplitude on the barrier action, for $|\varphi _0|<\pi /2$ the
tunneling to the states in the next cell, $\varphi _0\to \varphi _0\pm 2\pi $, can
be completely neglected, whatever the inductance of the system.
 
The supercurrent  through the normal part of the system is carried  by
a set of  Andreev levels formed by reflections at the ND boundaries \cite
{Andreev,Furusaki}. We will calculate it using the quasiclassical 
approach following from Eilenberger equations \cite{KulikOmelyanchouk}  
which is well suited to our problem. Here we
can consider "Andreev tubes"  along the quasiparticle trajectories in the
normal  part of the system\cite{Zagoskin,ZagoskinBarzykin,Fauchere}
each carrying the supercurrent density which in our case should be written 
as 
\begin{equation}
{\bf j}({\bf r},{\bf n}) = j_c {\bf n} \sum_{p=1}^{\infty} (-1)^p \frac{{\cal L}[{\bf r},{\bf n}]}{l_T} \frac{\sin p\varphi[{\cal L}[{\bf r},{\bf n}]]} {\sinh p{\cal L}[{\bf r},{\bf n}]/l_{T}}e^{-{\cal L}[{\bf r},{\bf n}]/l_{imp}}.  \label{1}
\end{equation}
Here the critical current density $j_c\sim (ev_F\Pi)/(\lambda_F S)$, $S$
and $\Pi$ being area and  half-perimeter of the system.  Equation (\ref{1})
is valid in the limit $\xi_0 \ll S/\Pi$, the  characteristic separation
between the superconducting electrodes and is thus  applicable to the case
of superconducting cuprates. (The opposite limit of large 
coherence length  is  considered in \cite{OmelyanchoukRezaei}.) 
The quantity ${\cal L}[{\bf r},{\bf n}]$ is the length of the
quasiclassical trajectory  linking two superconductors which  passes through
the point ${\bf r}$ in the direction ${\bf n}$; $\varphi[{\cal L}[{\bf r},
{\bf n}]]$ is the phase gain along this trajectory  (in this paper we do not
concern with the effects of current-induced field,  and therefore in the
absence of external fields this is simply the  phase difference between the
superconductors on the ends of the trajectory,  including the extra $\pi$ if 
${\cal L}[{\bf r},{\bf n}]$ connects the  lobe of the d-wave order
parameter with opposite signs\cite{Zagoskin}).  The normal metal coherence length $
l_T = v_F/2\pi k_BT$, and $l_{imp}$  takes into account effects of weak
elastic scattering by nonmagnetic  impurities (in the ballistic regime, by
definition, $l_{imp} \gg S/\Pi$). We have used standard approximation of
steplike behaviour of the order parameter at the ND boundary, and
neglected the own magnetic field of the supercurrent\cite{Svidzinskii}.

The total supercurrent density at a point {\bf r} is thus given by 
\begin{equation}
{\bf j}({\bf r}) = \int_0^{\pi} \frac{d\theta}{\pi} {\bf j}({\bf r},{\bf n}
(\theta)).
\end{equation}
Calculating the total current flowing in A, we find in the limit 
$l_{imp},l_T \to \infty$
\begin{equation}
I(\varphi) = \frac{2j_cW}{\pi}\left[ \frac{1+Z(\Omega)}{2}F(\varphi) +
 \frac{1-Z(\Omega)}{2}F(\varphi+\pi)\right],  \label{2}
\end{equation}
where $F(\varphi)$ is the $2\pi$-periodic sawtooth of unit amplitude, and
the imbalance factor $Z(\Omega)$ 
determines the equilibrium phase difference\cite{Zagoskin}
\begin{equation}
|\varphi_0| = \left|\frac{1-Z(\Omega)}{2}\right|\pi.
\end{equation}
In the setup of Fig.1a
\begin{equation}
|\varphi_0(\Omega)| = \frac{\sin|\Omega|}{\sqrt{2}}\pi.
\end{equation}
 
The current-phase dependence (\ref{2}) and corresponding Josephson energy 
$E_J(\varphi) = \frac{\hbar}{2e}\int^{\varphi} d\varphi I(\varphi)$ are plotted in Fig.2,
and current density distribution in the normal part of the system  is
shown in Fig.3. The vortex pattern is clearly seen.  The spontaneous flux 
in the system is 
\begin{equation}
\Phi_s \sim \kappa(\Omega) \frac{j_c\Pi^2}{c} \sim  \frac{\kappa(\Omega) eN_{\perp}v_F}{c}
\sim \frac{1}{137}\frac{\kappa(\Omega)}{\pi}\frac{v_F}{c} N_{\perp}\Phi_0,
\end{equation}
where $\kappa <1$ is a geometry-dependent attenuation factor (e.g. in SND junctions
$\kappa=0$ by symmetry if ND boundary is 
parallel to (100) or (010)
\cite{Huck,ZagoskinOshikawa}); $N_{\perp}\sim\Pi/\lambda_F$ is the number of 
transport modes in the system.

Let us make some estimates. Taking the size of the system $\sim 10^3$\AA, $v_F\sim 10^7$cm/s,
we find that $\Phi_s \sim \kappa\cdot 10^{-3}\Phi_0$. The magnetic moment
of the spontaneous current will be of order  
$m_s \sim \kappa(N_{\perp}e\Pi v_F)/c \sim \kappa\cdot 10^{5}\mu_B.$ 
The tunneling rate between the states is
$\Gamma\sim\omega_0\exp[-U(0)/\hbar\omega_0],$ where the frequency of oscillations 
near $\pm\varphi_0$ is $\omega_0\sim\sqrt{N_{\perp}\bar{\epsilon}\epsilon_Q}/\hbar$
and the height of the potential barrier $U(0)\propto (\varphi_0/2\pi)^2$
\cite{ZagoskinOshikawa}. 
Due to spatial Andreev quantization, there are no elementary
excitations in the normal part with the system with energies below 
$ \bar{\epsilon} \sim \hbar v_F/2\Pi \sim 10^{-15}$ erg. At temperatures below 
$\bar{T}=\bar{\epsilon}/k_B ~ 10$K  thermal excitations are frozen out, and
dissipation can only be due to interlevel transitions generated by   ac
Josephson voltage generated by phase fluctuations.
 Therefore it will be absent if
 \begin{equation}
2e<V_J> \sim \hbar\sqrt{<\dot{\varphi}^2>} \sim \hbar\omega_0 < \bar{\epsilon},
\end{equation}
which can be rewritten as $
\epsilon_Q < \bar{\epsilon}/N_{\perp}$ (where the charging energy $\epsilon_Q = 2e^2/C,$ and $C$ is the capacitance of 
the terminal A), 
 or
$\omega_0 < v_F/2\Pi. $
 The latter condition
is a physically
clear requirement 
that the quantum 
oscillations
of superconducting 
phase allow time 
for readjustment
of Andreev levels in the
system (which is indeed
$\sim 2\Pi/v_F$,
the time necessary 
for the electron
and Andreev reflected
hole to travel 
across the system). Otherwise the coherent transport through the
normal part of the system cannot be established, and dissipative
currents flow instead.   
The maximum value of 
$ \omega_0$ allowed by the above limitation is
$\omega_{\rm max}\sim 10^{12} {\rm s}^{-1}$.
(That is, the capacitance of the terminal A cannot be lower than $C_{\rm min}=2e^2N_{\perp}/\bar{\epsilon} \sim 10^{-11}$F.)
The corresponding tunneling rate is
\begin{equation} \Gamma_{\rm max} \sim \omega_{\rm max} \exp[-N_{\perp}(\varphi_0/2\pi)^2].\end{equation}
Therefore 
we would require $\varphi_0 \sim 0.2 \pi$ to have
tunneling rate in the 100 MHz region. 
 
For a system of integrated DND qubits of  Fig.1b the bulk
d-wave "bus" provides,   the possibility for
operations performed over all the qubits simultaneously by creating a supercurrent
flow along the bus. In particular, one can easily prepare the whole
register of qubits in the same (up or down) state. This is an attractive
property, e.g. for implementation of a quantum correction algorithm. 
If take the size of
a unit qubit with its periphery as $\sim 5\cdot 10^3$\AA, a 2D 100$\times$100-qubit
block will occupy only an area of 50$\times$50 $\mu$m$^2$, which is realistic to
keep below the dephasing length due to thermal excitations. 
 
 Application of quantum gates to individual qubits can be effected in various
ways, lifting the degeneracy between up/down states either by directly applying localized magnetic field to a qubit 
 (using a magnetic
scanning tip), by creating local supercurrents in the bus,
or by using laser beams with circular polarization. The entanglement of
the states of adjacent qubits is achieved simply by opening a
key between them for a certain time. 

In conclusion, we have suggested a new design for a solid state superconducting,
scalable qubit. Besides using the degeneracy of the ground state,
common to all D-D junctions, it strongly relies on unique properties of
clean DND or grain boundary junctions: tunability of the equilibrium
phase difference across the junction, and spontaneous  currents and fluxes in
equilibrium. The former allows to optimize the design in order to
achieve the fastest possible tunneling rate (which is vital in order to beat
the dephasing processes) and thus makes it easier to integrate qubits in a computer. The latter presents  an easier way to manipulate and read out
the state of a qubit.
Our estimates show that there is a real chance to create a working solid
state qubit along these lines using existing experimental possibilities.

{\em Acknowledgements}: I am grateful to  M. Beaudry, D. Bonn,
 S. Lacelle, P. Stamp and A.-M. Tremblay for helpful discussions and
critical comments,
and to the Dept. de physique, Universit\'e de Sherbrooke, 
for hospitality.
This research was supported in part by CIAR.

\newpage

\begin{figure}
\epsfxsize=4in
\epsfbox{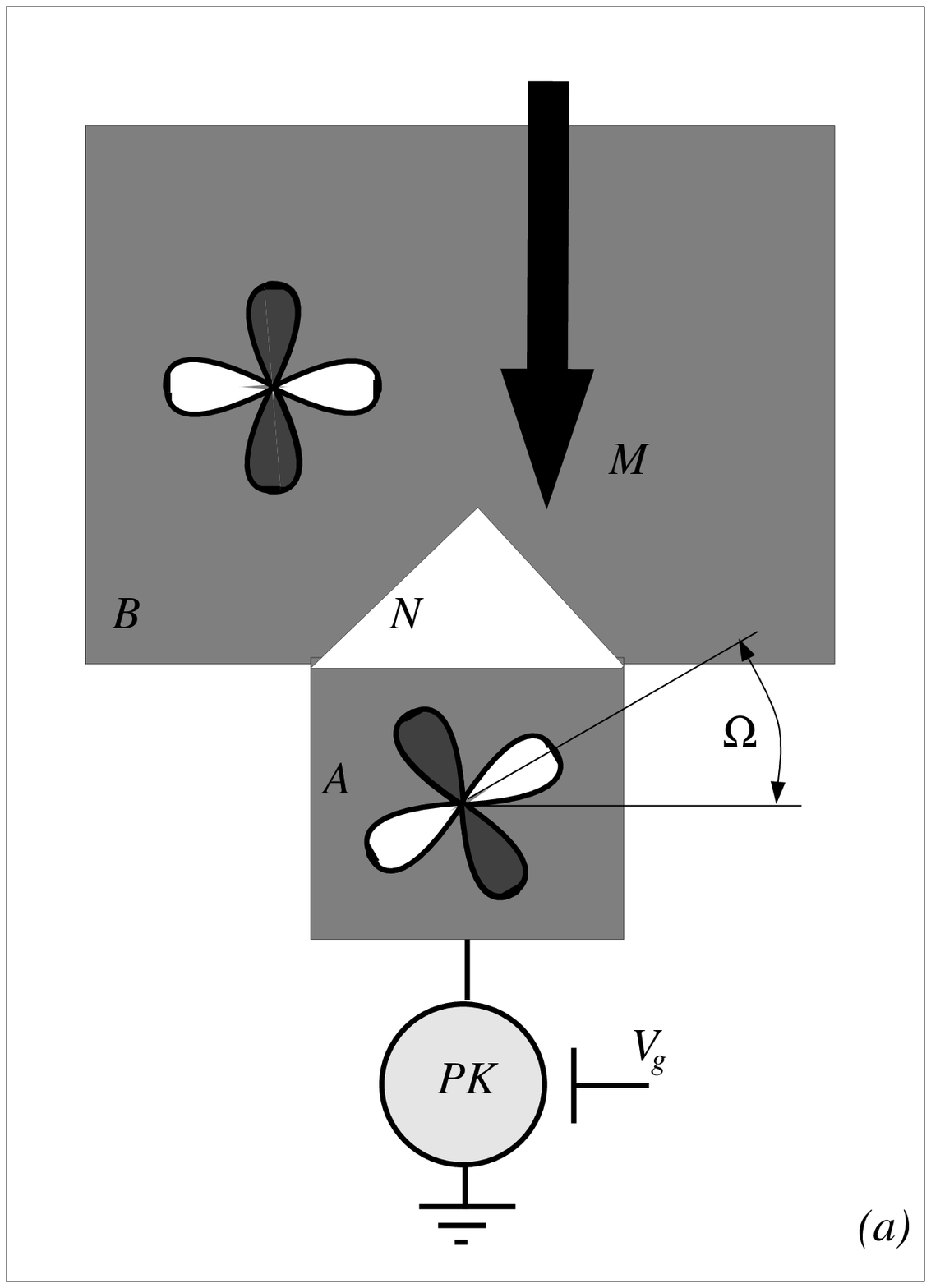}
\epsfbox{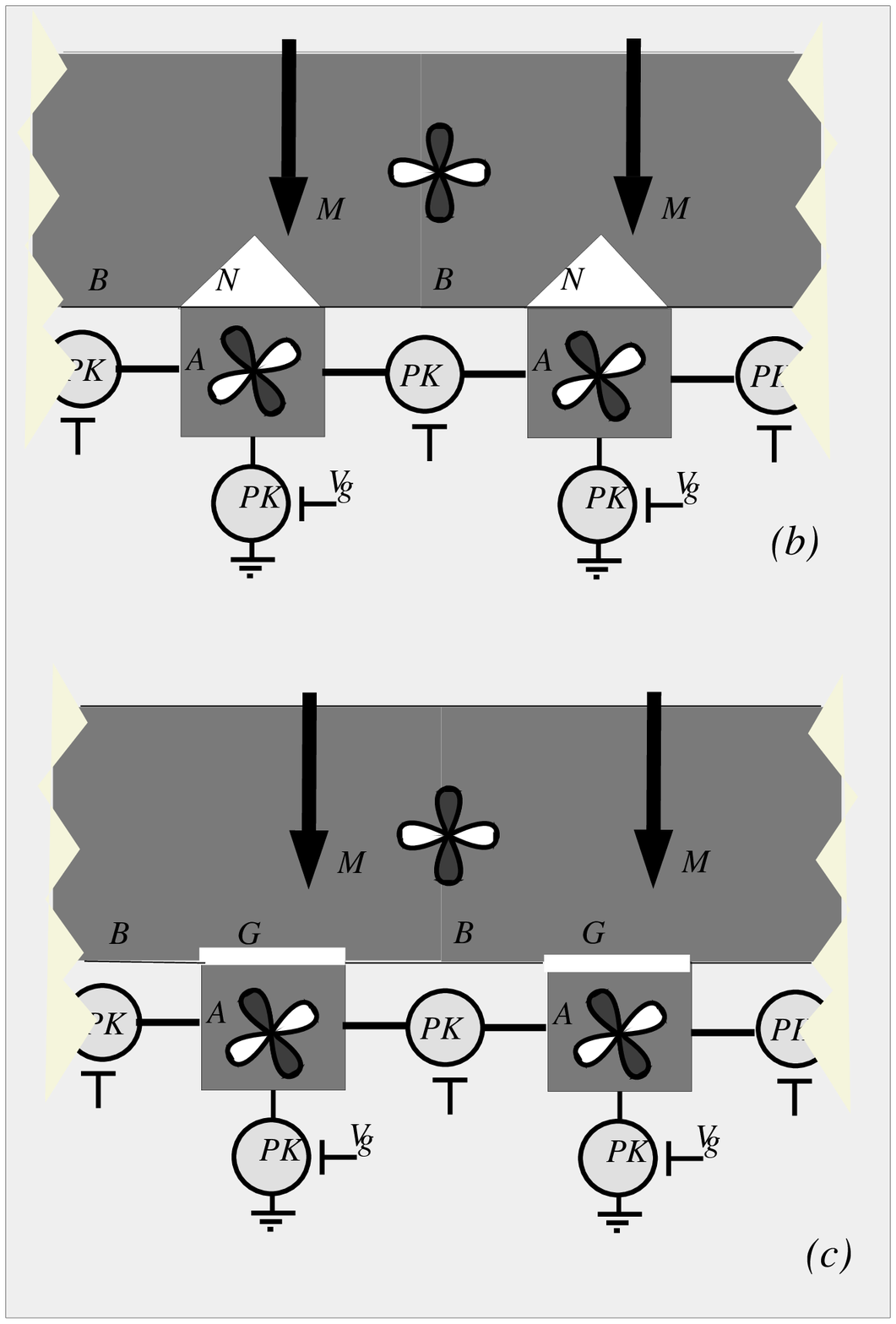}
\caption{(a) Superconducting DND qubit: A,B are d-wave superconductors, N  normal conductor,
PK  parity key, M  scanning tip, $\Omega$ the mismatch angle between the
lattices of A and B. The cut in B is here along $(110)$ and $(1\bar{1}0)$ directions. Positive lobes of d-wave order
parameter are shaded. (b) Multiqubit register. Terminal B plays the role of
the bus bar. (c) Version of (b) using grain boundary (G) junctions.}
\end{figure}

\begin{figure}
\epsfxsize=5in
\epsfbox{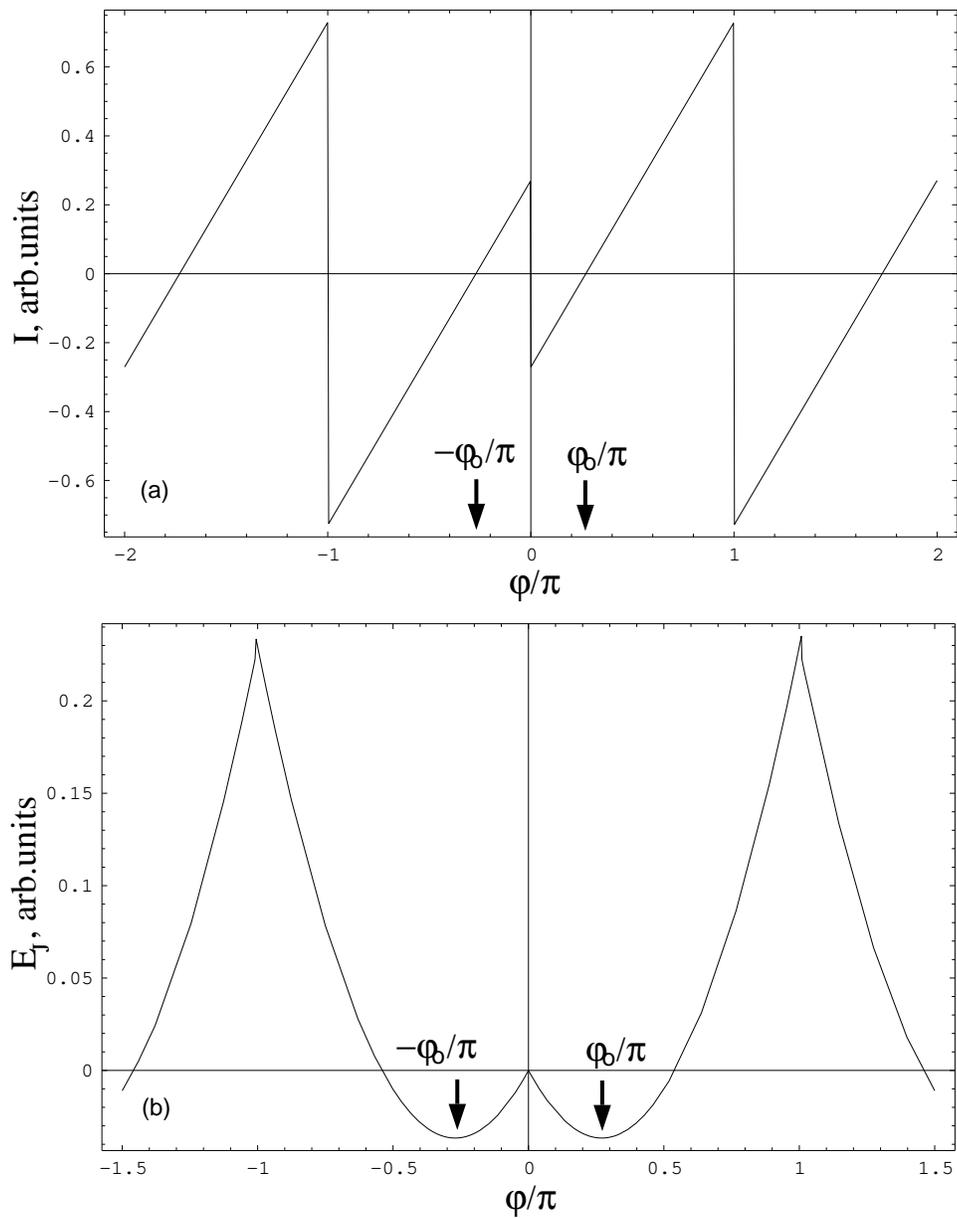}
\caption{ (a) Current-phase dependence in a DND junction of Fig.1 at 
$l_T, l_i\to\infty$. The
mismatch angle is $\Omega=\pi/8$. (b) Effective potential profile of the system. Minima at $\pm\varphi_0$ correspond
to "up" and "down" pseudospin states of a qubit.}
\end{figure}

\begin{figure}
\epsfxsize=5in
\epsfbox{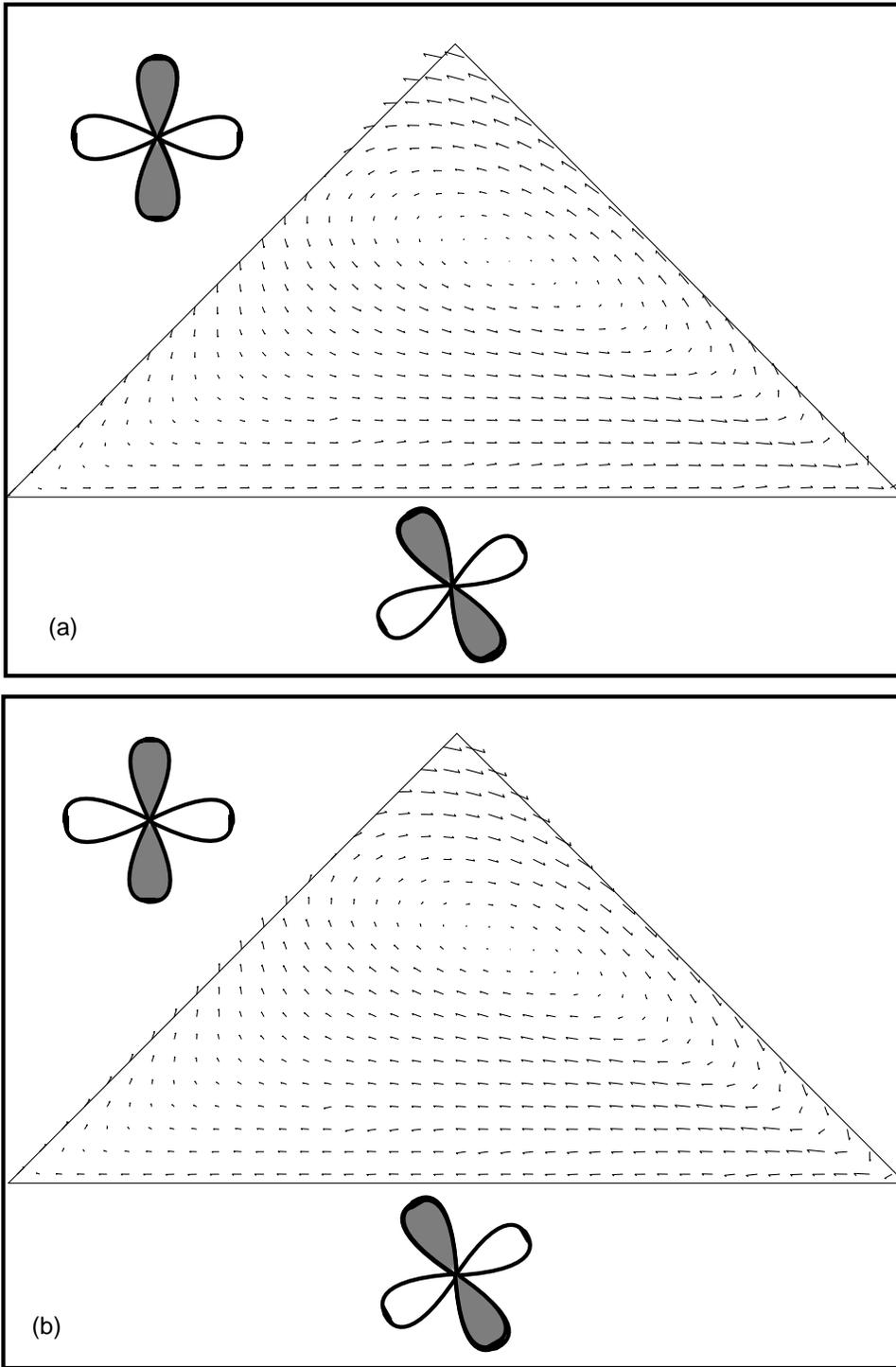}
\caption{Current distribution in the normal part of the system 
in degenerate 
equilibrium states: (a)$\Omega =  \pi/8,
\varphi = -\varphi_0 = -(\sin|\Omega|/2^{1/2})\pi \approx 0.27\pi; {\rm (b)} \Omega = \pi/8,
\varphi = \varphi_0 \approx 0.27\pi.$}
\end{figure}

\end{document}